# Diamondoid Structure of Polymeric Nitrogen at High Pressures

Xiaoli Wang[1,2], Yanchao Wang[3], Maosheng Miao[4,2], Xin Zhong[3], Jian Lv[3], Jianfu Li[1*], Li Chen[1], Chris J. Pickard[5], and Yanming Ma[3,2†]

[1]*Institute of Condensed Matter Physics, Linyi University, Linyi 276005, P. R. China*
[2]*Beijing Computational Science Research Center, Beijing, 100084, P. R. China*
[3]*State Key Lab of Superhard Materials, Jilin University, Changchun 130012, P. R. China*
[4]*Materials Research Lab, University of California Santa Barbara, CA 93110, USA*
[5]*Department of Physics and Astronomy, University College London, Gower Street, London WC1E 6BT, United Kingdom*

High-pressure polymeric structures of nitrogen have attracted great attention owing to their potential application as high-energy-density materials. We report the density functional structural prediction of the unexpected stabilization of a diamondoid (or $N_{10}$-cage) structure of polymeric nitrogen at high pressures. The structure adopts a highly symmetric body-centered cubic form with lattice sites occupied by $N_{10}$ tetracyclic cages, each of which consists of 10 atoms and is covalently bonded with its six next-nearest $N_{10}$ cages. The prediction of this diamondoid structure rules out the earlier proposed helical tunnel phase and demonstrates the high-order nature of polymeric nitrogen at extreme high pressures. Diamondoid nitrogen is a wide-gap insulator and energetically more favorable than the experimental cubic gauche and previously predicted layered *Pba*2 phases above 263 GPa, a pressure which is accessible to high pressure experiment.

Nitrogen exists abundantly in nature and stabilizes in the form of triply bonded diatomic molecules under ambient conditions. The nature of nitrogen is dramatically changed when subject to high pressure. The extremely strong triple N≡N bond dissociates into three weaker single N-N bonds at an unexpectedly modest pressure of 60 GPa [1-5]. In contrast, the molecular dissociation for oxygen is shown to happen at 1920 GPa [6-7] and for hydrogen at 500 GPa [8], even though these molecules have much weaker intra-molecular bonds. Because of the exceedingly large difference in energy between the single N-N and triple N≡N bonds, singly-bonded polymeric nitrogen has the potential to become an efficient high-energy-density material for energy storage, propellants, and explosives. The search for polymeric forms of nitrogen upon compression has therefore attracted great attention.

The early attempts [9-10] to describe singly-bonded polymeric structures of nitrogen focused on the simple cubic, black phosphorus (BP) and A7 (α-arsenic) structures, based on the known structures in group VA. Significant progress was made in 1992 by Mailhiot *et al.* [11] who predicted the *cubic gauche* (*cg*) structure. This remarkable structure was later found to be more energetically stable than all the previously proposed polymeric structures. Subsequently, numerous theoretical high-pressure structures of polymeric nitrogen were proposed, including *Cmcm* chain [12], relative cg, chaired web [13], $N_2$-$N_6$ [12], cis-trans chain [14], layered boat [15], and eight-member rings [16] structures. The use of sophisticated structural searching techniques that are unbiased by any known structure information have enabled us to predict new polymeric nitrogen structures, including the layered *P*-42$_1$/*m* and *Pba*2, and the helical tunnel $P2_12_12_1$ structures, which are energetically stable over appropriate pressure ranges[17-18]. These predicted structures rule out the BP structure and suggested a zero-temperature phase diagram of polymeric nitrogen at high pressures that follows the transition sequence of *cg* → *Pba*2 (188–320 GPa) → $P2_12_12_1$ (>320 GPa).

A great amount of experimental effort has been employed in synthesizing polymeric nitrogen, and evidence for non-molecular phases of nitrogen at high pressures has been reported by several groups [3-4, 19]. However, most samples are amorphous and likely to be mixtures of small clusters of non-molecular phases. In 2004, the crystalline form of singly-bonded polymeric nitrogen was synthesized by Eremets *et al* at high pressures (above 110 GPa) and high temperatures (above 2000 K) [20]. The obtained polymeric single crystal phase was found to be the long-sought *cg* structure, a finding that was confirmed by later independent experiments [21]. These experimental achievements motivate the investigation of other crystalline forms of polymeric nitrogen under higher pressures.

In this letter, we report an unusual stabilization of a diamondoid (or $N_{10}$-cage) structure of polymeric nitrogen at high pressures, firstly predicted by a CALYPSO structural search technique based on the particle swarm optimization (PSO) algorithm [22, 23] and then blindly confirmed in another independent structural search using *ab initio* random structure searching (AIRSS) [24, 25]. The prediction of the diamondoid structure rules out the previously proposed helical tunnel phase and substantially narrows the stable pressure range of the *Pba*2 phase, providing a significant step forward in the understanding the behavior of solid nitrogen and other nitrogen-related materials under extreme conditions.

The CALYPSO approach [22, 23] enables the global minimization of energy surfaces via *ab initio* total-energy calculations and the PSO algorithm. This method has been successful in correctly predicting high pressure structures for various systems [6, 26-28]. The underlying *ab initio* structural relaxations and electronic band structure calculations were performed in the frame work of density functional theory within generalized gradient approximation using Perdew-Burke-Ernzerhof functionals [29], as implemented in the VASP code [30]. Projector augmented wave [31] potentials are used to describe the ionic potentials. The cutoff energy (700 eV) for the expansion of the wave function into plane waves and Monkhorst-Pack [32] *k* meshes were chosen to ensure that all the enthalpy calculations are well converged to better than 1 meV/atom. The AIRSS structural search method has been described in details in Refs. 24 and 25. An ensemble of structures is chosen by first generating random unit cell translation vectors and renormalizing the resulting cell volumes to lie within some reasonable range. The atoms, or structural units, are then placed at random, possibly constrained by symmetry, positions and the cell shapes and atomic positions are relaxed at a fixed pressure to a minimum in the enthalpy. The structural optimizations and energy calculations for various structures have been performed by using CASTEP code [33].

CALYPSO structure predictions were performed for cells containing up to 80 N atoms at a pressure range from 100 to 800 GPa. Besides the earlier predicted *cg*, *Pba*2, *P*2$_1$2$_1$2$_1$ and *P*-421/*m* structures, our structural search identified a highly symmetric body-centered cubic (*bcc*) structure (20 atoms/cell, space group I-43m) as depicted in Fig. 1 (b). This cubic structure has a lattice parameter of *a* = 4.287 Å with N atoms sitting at two inequivalent crystallographic sites 12e (x, 0, 0) and 8c (y, y, y) with x = 0.3532 and y = 0.6745 at 300 GPa (3.939 Å$^3$/atom). Inspection reveals that the structure consists of identical N$_{10}$ cages at the *bcc* sites, each of which contains 10 N atoms forming a tetracyclic cage (Fig. 1a) and is covalently bonded with its six next-nearest neighboring N$_{10}$ cages. There is no covalent bonding between the nearest neighboring cages, although their center-to-center distances are smaller. Within each cage, there are 6 "bridge" N atoms (shaded in gray) and 4 "cage" N atoms (shaded in brown) as displayed in Fig. 1b. Although all the N atoms form three single N-N bonds with their neighbors, "bridge" and "cage" N atoms are inequivalent. Each "cage" N bonds equally with three "bridge" N atoms, forming a triangular pyramid with "cage" N sitting at the apex. In contrast, each "bridge" N bonds with two "cage" N atoms and one "bridge" N, forming a trigonal plane and the "bridge" N sits at the center. Analysis of electron localization functions (ELF) (Fig. 1c) suggests that the "cage" N atoms are *sp*$^3$ hybridized with lone pair lobes pointing opposite to the pyramid, whereas the "bridge" N atoms are in sp$^2$ hybridization with lone pair lobes ($p_z$) at both side of the triangular plane. As shown in Fig. 1(c), the lone pairs are arranged in a way that their lobes are avoiding each other by pointing to the perpendicular directions at neighboring N atoms. Following the prediction of the N$_{10}$ structure by CALYPSO, an independent structural search at 300 GPa was performed using the AIRSS method, with no prior knowledge of the structure beyond the number of atoms required to describe it. The $N_{10}$ structure was found and confirmed to be the most stable. Six new metastable structures were also identified [34] and included in the enthalpy curves (Fig. 2). Further structural design attempts based on using different combinations of $N_{10}$ units and applying symmetry did not yield any better structures.

The enthalpies (relative to the *cg* structure) of the $N_{10}$-cage structure together with other known and our newly predicted structures are plotted as a function of pressure in Fig. 2. It shows clearly that the *cg* structure is most stable up to 188 GPa, beyond which *Pba*2 structure is favored. The results are in excellent agreement with the previous calculations [12,17,35-36]. However, the prediction of the $N_{10}$-cage structure being stable at pressures higher than 263 GPa narrows the stability range of the Pba2 structure and rules out the $P2_12_12_1$ structure completely. Furthermore, a thorough CALYPSO structural search up to 800 GPa could not find any other structures that are more stable than the $N_{10}$-cage structure. Our current prediction has modified the earlier phase transition sequence into *cg* → *Pba*2 →$N_{10}$, and illustrates that polymeric nitrogen should become highly ordered at extreme pressures.

The electronic band structure calculations showed that the $N_{10}$-cage phase is an insulator. We plot the band structure at 300 GPa [Fig. 3 (a)] and the band gap as a function of pressure [Fig. 3 (b)]. Interestingly, the band gap of $N_{10}$-cage structure increases significantly with pressure, reaching 5.47 eV at 800 GPa (3.096Å$^3$/atom). Because density functional calculations usually lead to a considerable underestimation of the energy gap, the actual band gaps are expected to be much larger. The insulating state is the result of the complete localization of valence electrons, similar to the monatomic $O_4$ phase of oxygen [6-7]. The increase of the $N_{10}$-cage band gap with pressure is the result of the competition of two effects: (i) on one hand, the compression and the consequent shortening of the bond length will widen both valence and conduction bands, and therefore tends to reduce the gap; (ii) on the other hand, the stronger coupling of the $sp^2$ (or $sp^3$) orbitals at the neighboring N atoms will lower the energy of the bonding states (valence bands) and increase the energy of the anti-bonding states (conduction bands), therefore leads to an increase of the gap. In case of the $N_{10}$-cage structure, the gap increases with increasing pressure.

The stability of a structure cannot be determined exclusively by comparing enthalpy, since the structure might be subject to dynamic instabilities. Therefore, we calculated phonon dispersion curves for the $N_{10}$-cage phase using the supercell method [37]. No imaginary phonon frequencies are found in the pressure range of 250–1000 GPa in the whole Brillouin zone (Fig. 4), indicating that the $N_{10}$-cage structure is dynamically stable in this pressure range. The primitive cell of $N_{10}$-cage structure contains 10 atoms, giving 30 phonon branches. The calculated zone-center ($\Gamma$) phonon eigenvectors were used to deduce the symmetry labels of phonon modes. The group theory analysis shows that the 30 vibrational modes at the zone center has the irreducible representations $\Gamma_{N10} = 5T_2^{I+R} + 2A_1^R + 2E_1^R$, where the Raman active modes are labeled by superscript $R$, and infrared active modes are labeled with superscript $I$. Both infrared and Raman frequencies of the structure (zone-center phonons in Fig. 4) can provide useful information for future experiments to identify the novel $N_{10}$-cage phase as that was done in the case of the *cg* structure.

The $N_{10}$-cage structure reported here is highly unique and has not been seen in any other elements, including other group VA elements. Together with the facts that the earlier attempts to search for structures of polymeric nitrogen among known phases in group VA elements were not successful, it reveals a fundamental difference between N and other group VA elements in forming singly-bonded extended structures. N has a very small 1*s* core which is capable of forming very short bonds, in contrast to the much larger 2*p* core in P or 3*d* core in As. As a result, the low-pressure forms of BP, A7, and simple cubic phases adopted by P and As are not stable for nitrogen. Taking another view, if we consider a $N_{10}$ cage as a pseudoatom, the phase adopts indeed a bcc structure, which was adopted by P and As at very high pressures of 262 GPa and 110 GPa [38-40], respectively. In this regard, a $N_{10}$ cage or unit might be seen as a superatom of group VA. The novelty of $N_{10}$-cage structure lies in a compromise between atomic cores and atomic volumes tuned by external pressures.

Cage-like structures are rare in the elements and compounds. Besides the known B-cage in α-boron (and many other boron phases) [41-44], C-cages in $C_{60}$ or $C_{70}$ [45], and Si-cages in silicon clathrates (e.g., in $Ba_8Si_{46}$) [46], we have also found a H-cage in a hydrogen sodalite structure in $CaH_6$ [47]. C has the right number of electrons and forms conjugated π bonds. B is electron deficient and therefore forms multi-center bonds. The current discovery of the $N_{10}$-cage adds a surprising new member to the cage family, and extends the cage structures to group VA elements. However, compared with B and C, N has too many electrons and must develop a structure that is best in packing the lone pairs whose Coulomb repulsion excluded the stabilization of planar triangle, square, pentagon, or hexagon faces in the cages seen in other elements, and render the irregularly puckered-hexagon faces in $N_{10}$.

More strikingly, one single $N_{10}$ tetracyclic cage (Fig. 1a) can be derived by cutting out the eight vertex atoms of one single unit cell of cubic diamond and is the simplest possible nano-form of diamond, a so-called "diamondoid" [48], or by the removal of H from adamantane. In this regard, $N_{10}$ structure is a basic cage structure of diamond lattice, a *nitrogen diamond*. We tried larger cages cut from cubic or hexagonal diamond, no better structure was found. This is no surprise since the diamondoid structure is evidently denser. Intuitively, a cage structure is not the preferable choice for achieving best packing efficiency, but the diamondoid structure as a compromise adopts a mathematically perfect structure forming puckered tetracyclic cages to optimize the volume.

In summary, we have reported for the first time an unusual diamondoid structure for polymeric nitrogen under high pressures. Based on thorough structural searches and accurate density functional calculations, we predict that the diamondoid structure is stable in the pressure range > 263 GPa. The prediction of the diamondoid structure provides an unexpected example created by compression of a molecular solid and represents a significant step toward the understanding of the behavior of solid nitrogen and other nitrogen-related materials at extreme conditions. In view of the successful synthesis of the *cg* phase, it would be of great interest to experimentally synthesize the predicted diamondoid structure above 260 GPa and at a temperature higher than 2000 K. It will be necessary to consider the synthetic kinetics which have been found to be large in synthesis of *cg* phase.

## ACKNOWLEDGEMENTS

This work was supported by the National Natural Science Foundation of China (Nos. 11147007, 11025418, 91022029, 10974076) and Open Project of State Key Laboratory of Superhard Materials (Jilin University). MSM was supported by the ConvEne-IGERT Program (NSF-DGE0801627) and MRL (DMR-1121053). CJP is supported by the EPSRC. Part of calculations was performed in the High Performance Computing Center of Jilin University.

**Corresponding author:**

[†]mym@jlu.edu.cn

[*]lijianfu@lyu.edu.cn

**FIGURE CAPTIONS**

**FIG. 1.** (a) Side view of one $N_{10}$ cage to compare with the diamondoid. (b) Top view of $N_{10}$-cage structure. (c) The calculated ELF isosurface within one $N_{10}$ cage at ELF = 0.83.

**FIG. 2**. Enthalpy curves (relative to *cg* structure) of various polymeric structures as a function of pressure.

**FIG. 3.** (a) Band structure (left panel) and partial densities of states (DOS; right panel) at 300 GPa. (b) Band gaps calculated as a function of pressures for $N_{10}$-cage structure.

**FIG. 4**. Phonon dispersion curves for $N_{10}$-cage structure at 250 and 500 GPa, respectively.

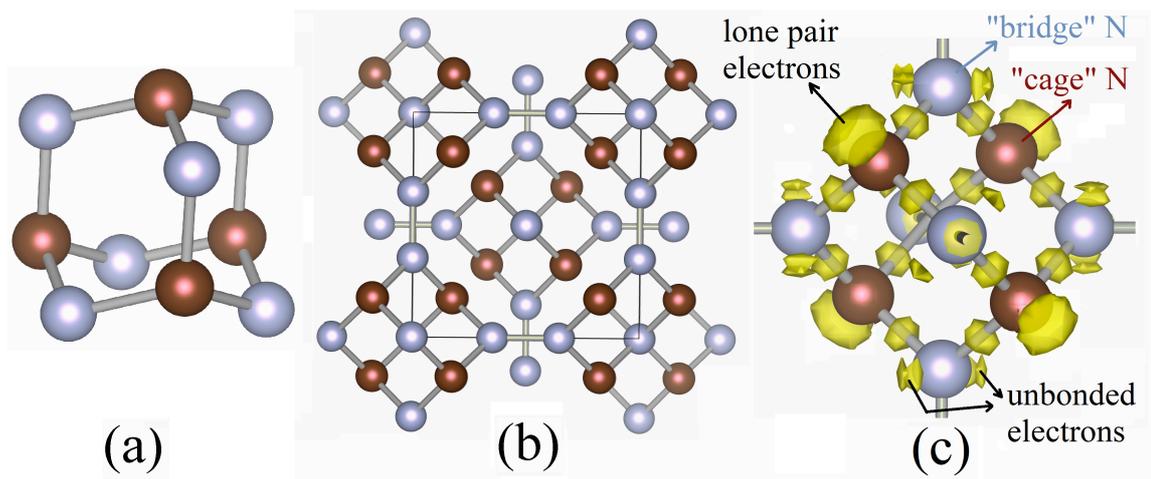

**FIG. 1**

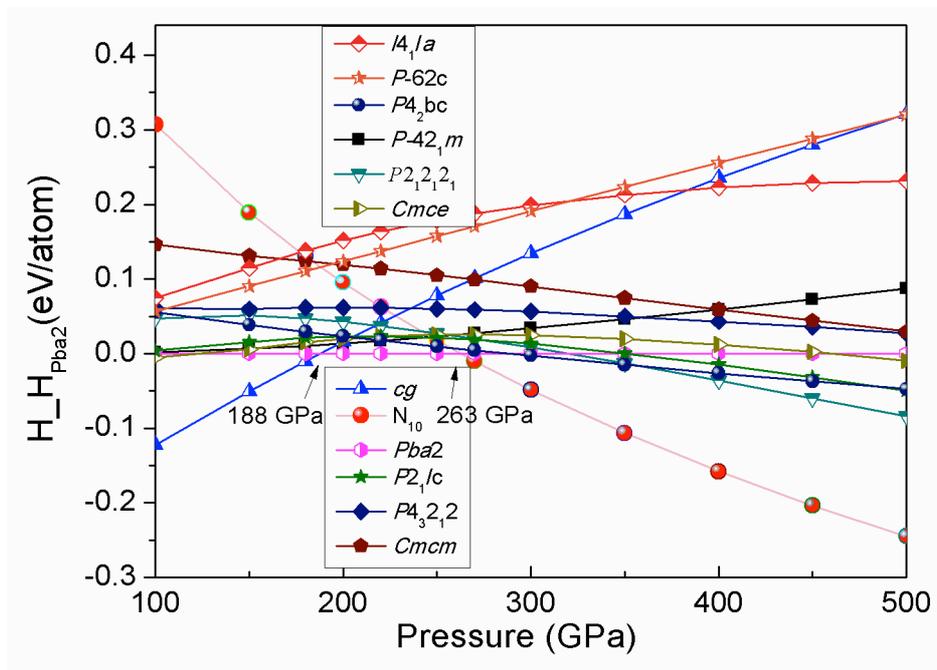

**FIG. 2**

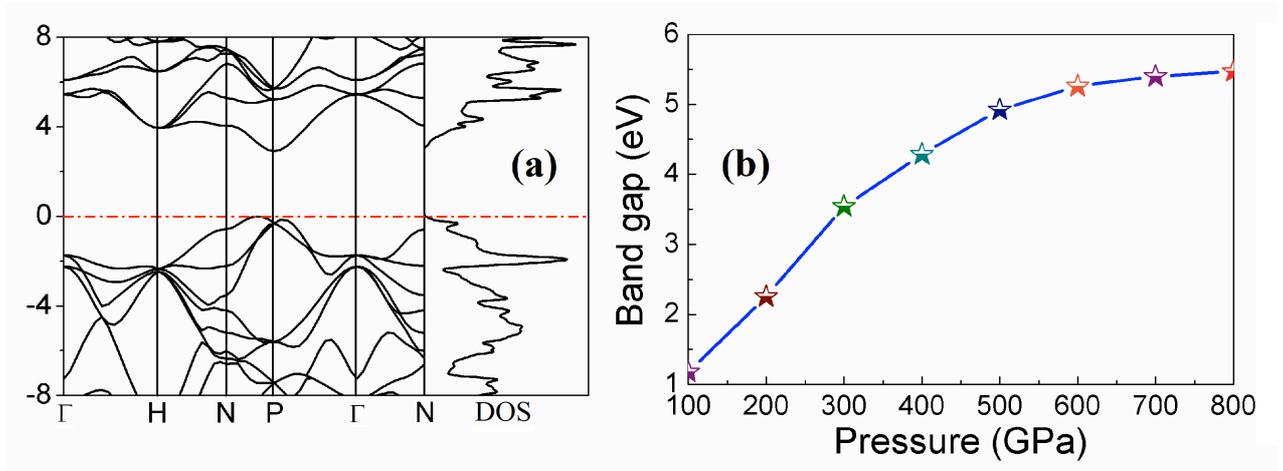

**FIG. 3**

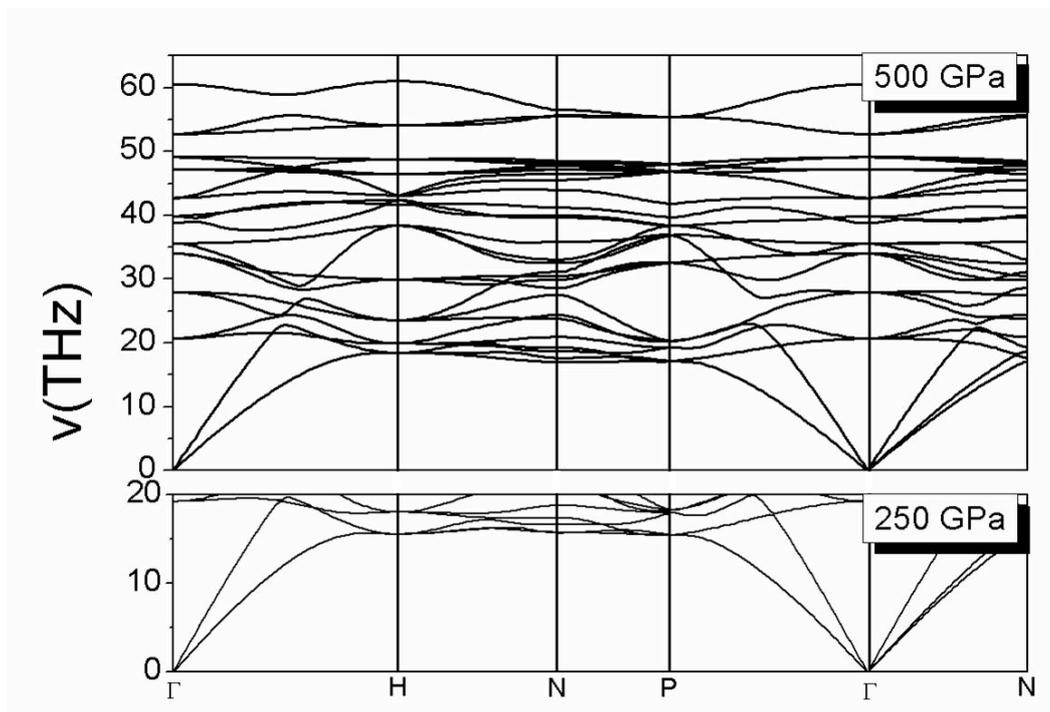

**FIG. 4**